\documentclass[twocolumn,showpacs,prl]{revtex4}

\begin{document}

\title {Observation of spin-wave characteristics in the two-dimensional
ferromagnetic ordering of in-plane spins}
\author {M. K. Mukhopadhyay$^1$, M. K. Sanyal$^1$, T. Sakakibara$^2$,
V. Leiner$^3$, R. M. Dalgliesh$^4$, and S. Langridge$^4$}
\affiliation {$^1$Surface Physics Division, Saha Institute of Nuclear Physics,
1/AF, Bidhannagar, Kolkata 700 064, India.\\
$^2$Institute for Solid State Physics, University of Tokyo,
Kashiwanoha, Kashiwa, Chiba 277-8581, Japan.\\
$^3$Lehrstuhl f\"{u}r Experimentalphysik,
Ruhr-Universit\"{a}t Bochum, D44780 Bochum, Germany.\\
$^4$ISIS, Rutherford Appleton Laboratory, Chilton, Didcot,
Oxfordshire OX11 0QX, UK}
\date{\today}

\begin{abstract}

The role of dipolar interactions and anisotropy are important to
obtain, otherwise forbidden, ferromagnetic ordering at finite
temperature for ions arranged in two-dimensional (2D) arrays
(monolayers). Here we demonstrate that conventional low
temperature magnetometry and polarized neutron scattering
measurements can be performed to study ferromagnetic ordering of
in-plane spins in 2D systems using a multilayer stack of
non-interacting monolayers of gadolinium ions. The spontaneous
magnetization is absent in the heterogenous magnetic phase
observed here and the saturation value of the net magnetization
was found to depend on the applied magnetic field. The net
magnetization rises exponentially with lowering temperature and
then reaches saturation following a $T\ln(\beta T)$ dependence.
These findings verify predictions of the spin-wave theory of 2D
in-plane spin system with ferromagnetic interaction and will
initiate further theoretical development.

\pacs {75.70.Ak, 75.60.Ej, 75.50.Xx}
\end{abstract}
\maketitle

Ferromagnetic materials confined in ultra-thin films and
multilayered structures are being studied extensively for the
development of high-density magnetic data storage devices and to
refine our basic knowledge in low-dimensional physics
\cite{aharbook,handbook,merminwagner,brunoprl01}. Recent advances
in growth techniques such as molecular beam epitaxy and
magnetization ($M$) measurement techniques based on the
magneto-optical Kerr effect have enabled us to measure small
magnetic signals as a function of applied magnetic field ($H$) and
temperature ($T$) even from one atomic monolayer of a
ferromagnetic material and a wide range of ordering effects has
been observed \cite{elmer,bellrmp,brunoprb,kashuba,maitinature,
macisaac}. These measurements have also demonstrated the existence
of a spontaneous magnetization and have revealed hysteresis curves
in two- \cite{elmer} and one-dimensional \cite{maitinature}
systems, where magnetic ions are arranged in a grid or in a line
within a monolayer. A recently generalized theorem
\cite{merminwagner,brunoprl01} showed, following spin-wave theory,
that long-range ferromagnetic order and hence spontaneous
magnetization cannot exist at finite temperature in a
two-dimensional systems provided spin-spin interactions are
isotropic and short range. A theoretical formalism
\cite{brunoprb,kashuba,maleev,brunomat} and computer simulations
\cite{bellrmp,macisaac} have been developed to include anisotropy
and dipolar interactions to explain the apparent contradiction
between theory and experiment in low-dimensional systems. A 2D
array of magnetic ions with lattice parameter `$a$' of spins $S$
can be described by a Hamiltonian,
\begin{equation}
{\cal H}={\cal H}_{ex}+{\cal H}_d+{\cal H}_k
\end{equation}
The strength of the three terms arise from exchange, dipolar and
magneto-crystalline anisotropic interactions respectively, and
have been approximated by expressing \cite{bellrmp,brunoprb} these
terms in equivalent magnetic field units as,
\begin{equation}
2\mu_B H_{ex}=JS\, ,~2\mu_B H_d=4\pi\alpha gS\, ,~2\mu_B H_k=6KS
\end{equation}
In the above expression $\alpha(\sim 1)$ depends on the lattice
type and $g$ is equal \cite{bellrmp} to $(2\mu_B)^2/a^3$. $K$ is
the anisotropy constant. The magnetization reduction due to
thermally activated spin waves was calculated with this
Hamiltonian and one obtains a non-zero temperature for long-range
ferromagnetic ordering as a gap of width $\Delta_z=2\mu_B
H_k^{\mathrm{eff}}$ opens up at the bottom of the spin wave
spectrum for an easy-magnetization axis ($z$) perpendicular to the
film plane. The easy-magnetization axis is determined by the sign
of the effective anisotropy field $(H_k^{\mathrm{eff}}=H_k-H_d)$
defined \cite{brunoprb} by
\begin{equation}
H_k^{\mathrm{eff}}=\frac{1}{2\mu_B}(6K_{\mathrm{eff}}S)
\hspace{0.1in}
\mathrm{with}\hspace{0.1in}K_{\mathrm{eff}}=K-\frac{2\pi\alpha
g}{3}\nonumber
\end{equation}
This explains the observation of hysteresis curves in a monolayer
with spins oriented normal to the surface
\cite{elmer,bellrmp,maitinature}. However, the situation is quite
different for spins oriented in an in-plane direction with
$H_k^{\mathrm{eff}}<0$ as the spin-wave spectrum remains gapless.
The long-range character of the dipole interactions was found
\cite{maleev} to be responsible for creating a pseudo-gap
$\Delta_{xy}=(\pi Sg/2) \sqrt{(6|K_{\mathrm{eff}}|/J)}$ in the
spin-wave spectrum that may give rise to long-range ferromagnetic
order in 2D in-plane spins provided
$|K_{\mathrm{eff}}|>K_c=\pi^2g^2/(6J)$. The temperature dependence
of the magnetization $M(T)$ above a transition temperature
$T_c\,(=6S|K_{\mathrm{eff}}|/K_B)$ takes the same form
\begin{equation}
M(T)=M_0[1-A T \ln(\beta T)]
\end{equation}
for the ordering of in-plane spins (with
$\beta=K_B/\Delta_{xy}=2\sqrt{6|K_{\mathrm{eff}}|J}/(\pi g T_c))$
and out-of-plane spins (with $\beta=K_B/\Delta_z=1/T_c$). Here
$A=K_B/(4\pi JS^2)$ and $M_0$ is the saturation value of the net
magnetization that depends on the field applied to carry out
measurements \cite{brunomat}. Below $T_c$ spin wave theory
predicts \cite{brunoprb} an enhancement of $M(T)$ as
$[1-A\,\exp\left(-1/(\beta T)\right)]$ and $[1-CT^{\nu}]$ for
out-of-plane and in-plane ordering respectively where $C$ depends
on $\Delta_{xy}$ and $\nu$ expected \cite{bellrmp, brunoprb,
maleev} to be $3/2$. For $0<|K_{\mathrm{eff}}|<\pi^2 g^2/(6J)$
in-plane spins can not stabilize in a homogeneous phase as the
magneto-crystalline anisotropy becomes large enough to pull some
of the spins in the out-of-plane direction and create a ripple
like instability \cite{bellrmp, brunoprb}. It is known that both
the average magnetization $M(T)$ as well as the initial
susceptibility $\chi(T)$ is proportional to the physical extent of
the ordered phase $l^{\star}$ that minimizes the zero-field energy
\cite{bellrmp} and can be written as
\begin{equation}
M\propto Hl^{\star} \hspace{0.1in}\mathrm{and}\hspace{0.1in}
\chi\propto l^{\star}\hspace{0.1in}\mathrm{with}\hspace{0.1in}
l^{\star}\propto \exp(-\gamma T)
\end{equation}
It is expected that at low enough temperature, $l^{\star}$ reaches
saturation either because $l^{\star}$ becomes comparable to the
sample size or due to a freezing of the ripple walls. The net
magnetization $M(T)$ of the ordered domains then should follow the
spin-wave prediction (Eq.\,(4)) to reach saturation.

Here we present the verification of this theoretical prediction of
2D ferromagnetic ordering of in-plane spins using
Langmuir-Blodgett (LB) films of gadolinium stearate. The presence
of a large multilayer stack of non-interacting monolayers of
gadolinium has enabled us to carry out conventional quantitative
magnetization measurements at sub-Kelvin temperatures. We could
also use polarized neutron scattering measurements \cite{shull,
felcherrapid}, to show that the ordered phase of the in-plane
spins is inhomogeneous and that the monolayers remain uncorrelated
even when the net magnetization reaches a saturation value.

In the metal-organic structure formed by LB techniques
\cite{prb02, joyphysrep, prb03}, gadolinium ions are separated by
approximately 5\AA~within a monolayer to form a distorted
hexagonal 2D lattice and the monolayers are separated from each
other by 49\AA~by organic chains (Fig.\,1). Films having 50 such
monolayers of gadolinium ions were deposited on 1 mm thick Si(001)
substrates and characterized using x-ray reflectivity technique,
as discussed earlier \cite{prb03}.

Neutron reflection measurements (refer Fig.\,2) were carried out
on the CRISP reflectometer at the Rutherford Appleton Laboratory
(RAL), UK using a cold, polychromatic neutron beam
\cite{penfoldcondmatt, langridge} and at the ADAM beamline
\cite{zabelphysicaB, leinerprl} of the Institut Laue-Langevin
(ILL), Grenoble, France, using a monochromatic cold neutron beam.
In Fig.\,2(b) we have shown spin-polarized neutron reflectivity
data taken at 2K by applying a magnetic field of 13 kOe along an
in-plane direction (refer Fig.\,1). It is known that, if the
polarization of the neutrons defined to be parallel (+) or
antiparallel (-) to the applied field direction (along +y axis)
\cite{shull, felcherrapid, kepa}, the intensity of a multilayer
Bragg peak in this geometry increases with effective scattering
length $b_{\mathrm{eff}}=b\,\pm\,A\mu_y$ with $A = 0.2695 \times
10^{-4}$ \AA /$\mu_B$. In the left inset of Fig.\,2(b) we have
shown intensity profiles of parallel (+) and anti parallel (-)
incident neutrons at the first peak position. The average of (+)
and (-) profiles represent the non-magnetic contribution to the
reflectivity. Systematic analysis of all these profiles provide us
the value of $\mu_y$, the component of average moment per
gadolinium ion along +$y$ direction. The obtained values of
$\mu_y$ as a function of $H$ at 4.2K and at 1.75K using the CRISP
and ADAM spectrometers respectively are shown in Fig.\,2(a) along
with data obtained from the magnetization measurements
\cite{prb03} at 2K and 5K. Results of these two independent
measurements show that the obtained average saturated moment per
gadolinium ion is much less than the expected 7$\mu_B$ and
confirms the existence of a heterogeneous phase. In the right
inset of Fig.\,2(b) we have shown the transverse diffuse neutron
scattering intensity profile at the first Bragg peak. The
hyper-geometric line shape profile confirms that the in-plane
correlation is logarithmic in nature and that the interfaces are
conformal \cite{diffuseprl, diffusedata}. It should be noted that
unlike in x-ray measurements the scattering here originates
primarily from the metal heads. The line shape and the associated
parameters were found to be independent of $T$, $H$ and hence
exhibit the absence of conformality in the magnetic correlations
between interfaces. This again confirms that the LB films
represent a collection of isolated 2D spin-membranes of gadolinium
ions .

Now we present the results of conventional magnetization
measurements carried out at sub-Kelvin temperature to understand
the nature of the ordering. These measurements were performed in a
conventional way by measuring forces exerted on a sample situated
in a spatially varying magnetic field with a Faraday balance
\cite{sakajjap}. In Fig.\,2(a) we have shown $M$ vs. $H$ data
taken at 100mK and 500mK temperature in an in-plane ($+y$)
direction. This data reconfirms the absence of hysteresis and
remanence ($M=0$ at $H=0$). The saturation value of the net
magnetization at 100mK and 500mK found to be $12.7\times10^{-6}$
emu/mm$^2 \cong 5.4 \mu_B$/ Gd atom; much lower than the expected
$7.0\,\mu_B$/Gd atom for a homogeneous ferromagnetic phase.

In Fig.\,3(a) we have shown the magnetization data taken with
different fields as a function of temperature. At higher
temperatures these data were fitted with Eq.\,(5), and the
expected exponential dependence is also observed in the
magnetization data extracted from our neutron reflectivity
measurements (inset of Fig.\,3(a)). The values of $\gamma$
obtained from fitting were found to increase with the reduction of
$H$ and at 0.25 kOe it is found to be $2.162\, K^{-1}$. It is also
observed that at a lower field, the magnetization at a fixed
temperature is nearly proportional to the applied field
($5.03\times10^{-7}$ at 0.25 kOe and $1.29\times10^{-6}$
emu/mm$^2$ at 0.5 kOe and at a temperature 0.9\,K) as predicted
for inhomogeneous striped phases (refer Eq.\,(5)). The amount of
majority phase grows exponentially as we lower the temperature for
each applied field but below a certain temperature this growth
stops as the walls of the majority phase freeze. Below this
temperature measured data do not follow Eq.\,(5) and the net
magnetization of the majority phase $M(T)$ increases with lowering
temperature following the thermally activated spin waves as given
in Eq.\,(4). We fitted all the data with Eq.\,(4) and obtained
$M_0$ values as $0.9\times10^{-6}, 1.6\times10^{-6},
3.2\times10^{-6}, 4.5\times10^{-6}, 7.9\times10^{-6},
9.5\times10^{-6}$ emu/mm$^2$ with 0.15, 0.25, 0.5, 1.0, 2.5, 5.0
kOe magnetic fields respectively. These saturation values of net
magnetization indicate that the percentage of the ferromagnetic
phase is increasing from 7.1\% to 74.8\% as we approach the
maximum saturation value of net magnetization obtained of
$12.7\times10^{-6}$ emu/mm$_2$ ($\cong 5.4~\mu_B$/Gd atom) as
shown in Fig.\,2(a). We extracted the value of exchange $J$ as
$8.76\times10^{-19}$ erg (or $H_{ex}$= 0.165 kOe) from the fitted
value of $A(=1.02\,K^{-1})$ for the 0.25 kOe data. We obtained
$\beta$ as 3.4 for all the data and hence $|K_{\mathrm{eff}}|$ was
calculated to be $1.7\times10^{-19}$ erg (or
$H_k^{\mathrm{eff}}=0.19$ kOe) for 0.25 kOe data giving $T_c$=26
mK. In this calculation $g$ was $6.88\times10^{-18}$ erg, assuming
that one gadolinium atom occupies 2.5 \AA$\times$20 \AA$^2$, as
obtained from neutron and x-ray analysis (refer Fig.\,1). This
proves that we are dealing with an inhomogeneous phase as
$0<|K_{\mathrm{eff}}|<K_c\,(=8.89\times10^{-17}\,erg)$. Although
Eq.\,(4) describes the temperature dependence of the magnetization
for ferromagnetic ordering of both in-plane and out-of-plane
spins, the argument of the logarithmic function can become less
than 1 only for in-plane ordering. Unusually low values of
$H_{ex}$ and $H_k^{\mathrm{eff}}$ with a rather large value of
$H_d(=16.3\,kOe)$ make $\beta T<1$ even for $T>T_c$ -- this
situation has not been reported earlier to the best of our
knowledge. It is interesting to note that all the magnetization
data shown in Fig.\,3(a) attains respected saturation values $M_0$
at temperature $T_0=1/\beta\, (\simeq 0.29\,\mathrm{K})$ and a
maximum magnetization at temperature $T_m=1/(e\beta)\, (\simeq
0.108\,\mathrm{K})$. Our experimental uncertainties below 100 mK
prohibit us from commenting on any reduction of magnetization
below this $T_m$ but Eq.\,(4) with $\beta T<1$, all the data is
fitted quite well. Further theoretical development is required to
understand the thermal activation of spin-waves with in-plane
ordering especially as we approach below $T_c(=26\,mK)$.

In Fig.\,3(b) we have shown zero-field-cooled (ZFC) and
field-cooled (FC) magnetization data taken with 0.15 kOe and 0.5
kOe field along with the fitted curve from Eq.\,(4). We observe a
blocking temperature ($T_b$) of 125 mK below which there is
branching in both the ZFC and FC data. This result reconfirms that
the observed ferromagnetic ordering requires an external field to
stabilize and such a low $T_b$ indicates the existence of very
small spin clusters in the ZFC phase. Application of the field
lowers the activation energy required for the randomly oriented
domains to increase the number of spins in the ferromagnetically
ordered majority phase and increases the activation energy of the
reverse transition \cite{bellrmp}. As a result we do not observe
$T_b$ in the measured temperature range for 0.5 kOe.

In conclusion, we have demonstrated that polarized neutron
scattering and conventional magnetization measurements can be used
to study 2D ferromagnetic ordering of in-plane spins using a stack
of magnetically uncorrelated spin-membranes formed with gadolinium
stearate LB film. The in-plane ordering observed here shows that a
spontaneous magnetization is absent even at 100 mK and saturation
value of the net magnetization increases with a lowering in
temperature. The magnetization is found to increase exponentially
with a lowering in temperature due to the exponential increase of
the physical extent of the ferromagnetic domains in the
heterogeneous phase. The ferromagnetic domains ultimately saturate
following $T\,\ln(\beta T)$: characteristic of thermally activated
spin-waves and are found to be valid for even $\beta T\leq 1$. We
believe these experimental results will initiate further
theoretical development.

\begin{figure}
{\bf Figure captions:} \caption{(a) Schematic diagram of the
out-of-plane and in-plane structure of the gadolinium stearate
Langmuir Blodgett film is shown with the scattering geometry
employed for the polarized neutron reflectivity measurements.
$x-z$ plane is the scattering plane and the magnetic field is
applied along the $+y$ direction. $\lambda$ and $\alpha$ are the
wavelength of the radiation and angle of incidence respectively.}

\caption{(a) In-plane Magnetization curves obtained as a function
of the field ($H$) using neutron reflectivity measured at 4.2K
(diamond) and 1.75K (star) compared with conventional
magnetization data\cite{prb03} measured at 2K (down-triangle) and
5K (up-triangle). Solid lines are the fits with a modified
Brillouin function \cite{prb03}. Magnetization measured at 100mK
and 500mK is shown for the 1st (symbols) and 2nd (line) cycle of
the hysteresis loop. (b) Neutron reflectivity data (symbols) at
$H$ = 13 kOe and at $T$=2K for the neutron spin along (+) and
opposite (-) to the magnetic field direction with the
corresponding fit (line). In the left inset the first Bragg peak
is shown in (+) and (-) channels in an expanded scale. Right
inset: Transverse diffuse neutron scattering profiles (symbols)
measured at 2K with unpolarized and polarized neutron beams. The
solid line is a fitted hypergeometric curve as described in the
text.}

\caption{(a) Sub-Kelvin magnetization results with various applied
fields (symbols) fitted with Eq.\,(4) (Black line) and with
Eq.\,(5) (wine coloured dashed lines). Dotted lines indicate the
temperatures $T_m$ and $T_0$ (refer to text). Inset shows the
magnetization obtained from neutron measurements as a function of
temperature (symbols) and fit with Eq.\,(5) (line). (b) ZFC (green
circles) and FC (blue stars) along with the fit (line) for FC
measurements.}

\end{figure}


\begin{thebibliography}{30}
\bibitem{diffusedata}The measured roughness and `true'
roughness \cite{diffuseprl} of the interfaces are calculated to be
2.3 \AA~ and 5.4 \AA~ respectively and we get an effective surface
tension equal to 10 mN/m.
\end{thebibliography}
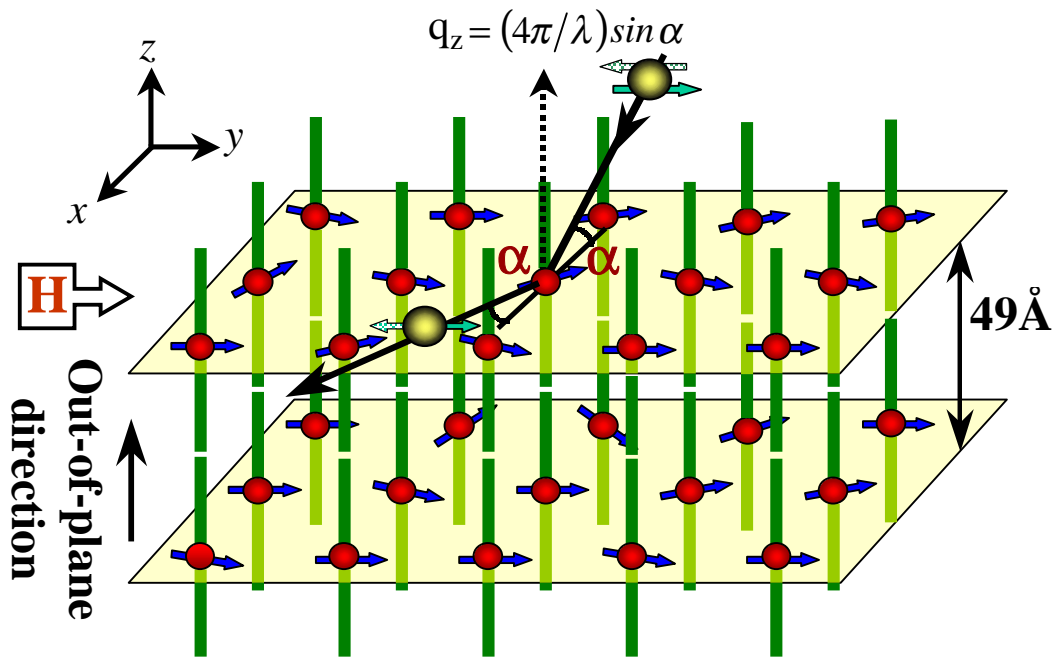
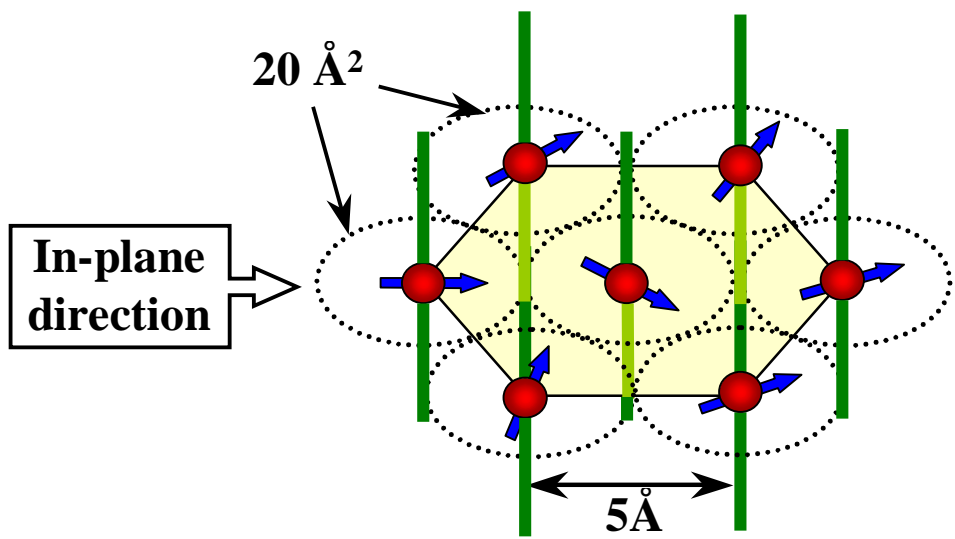
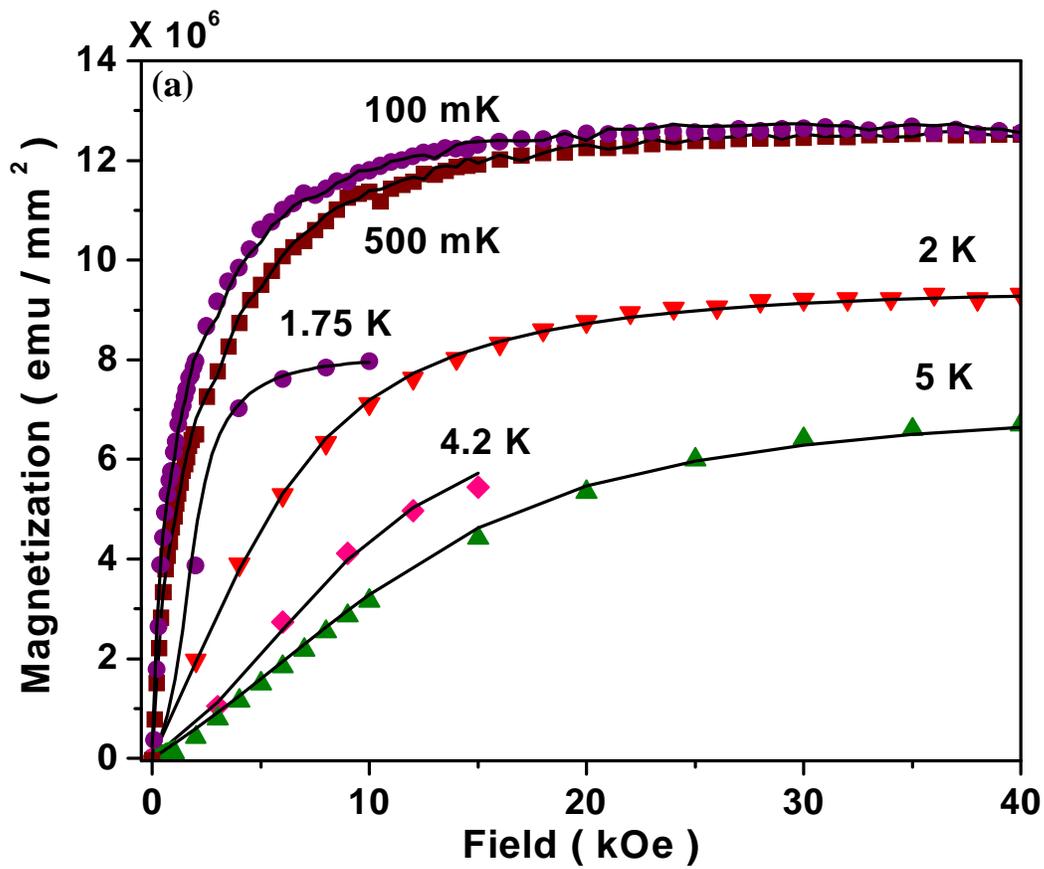
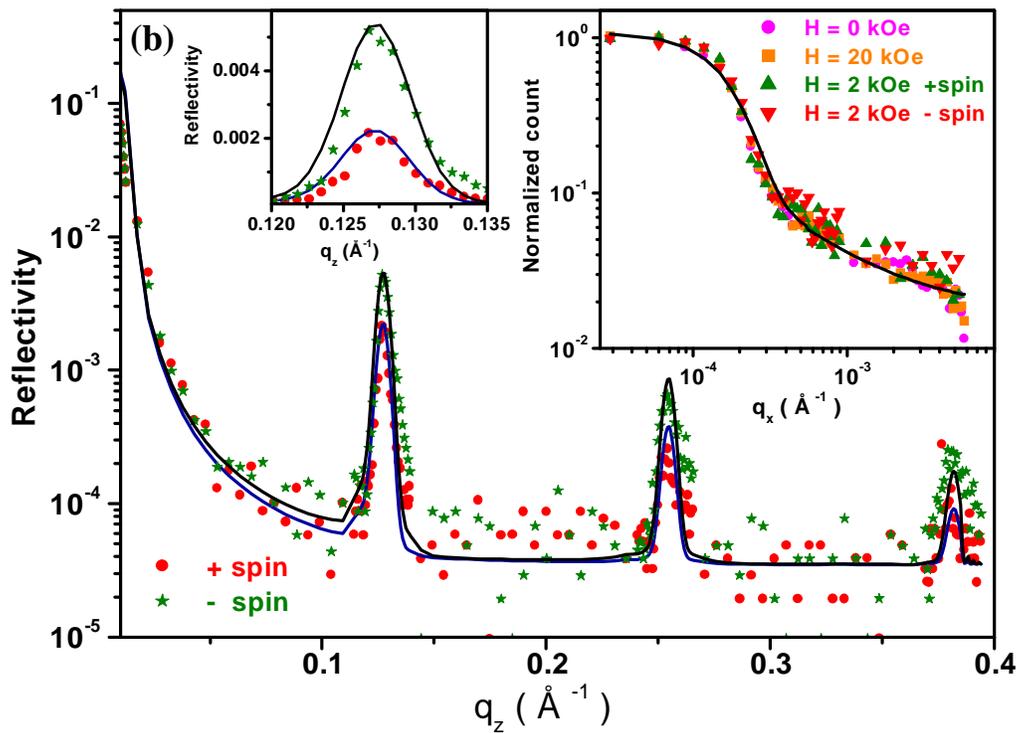
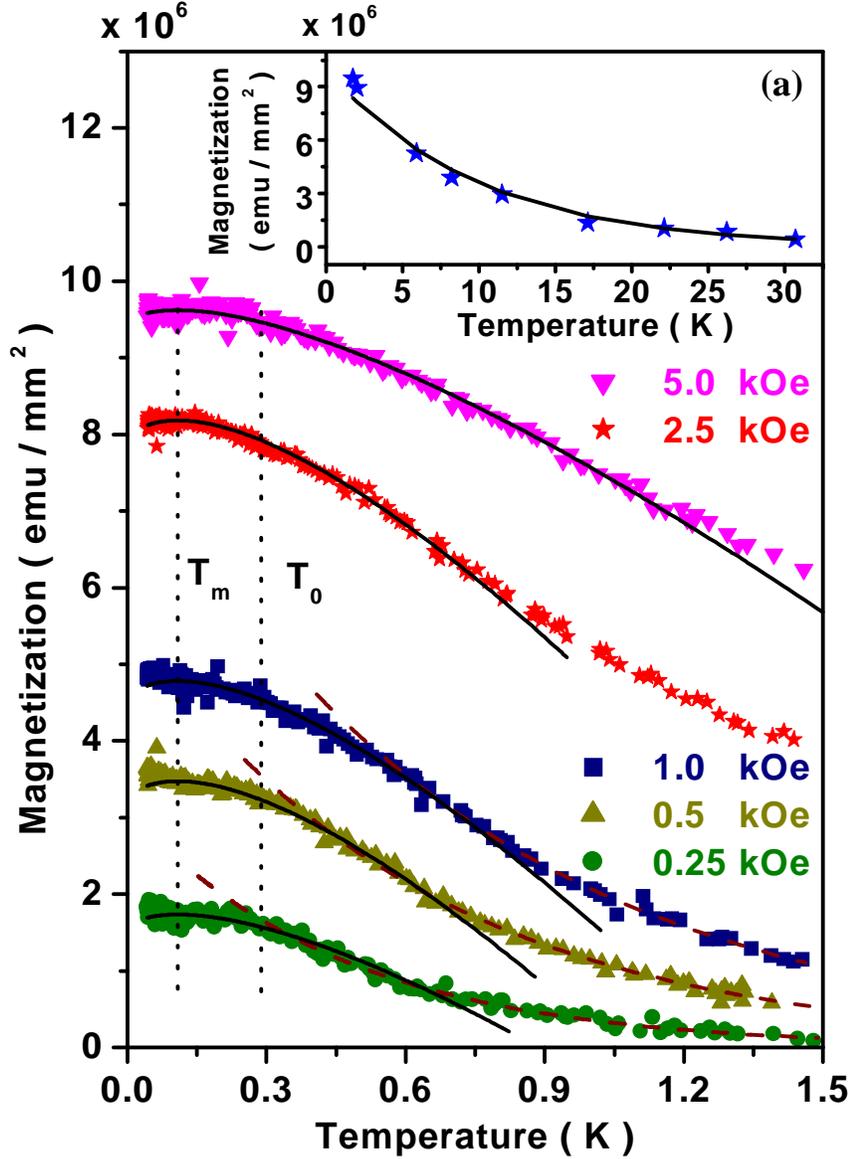
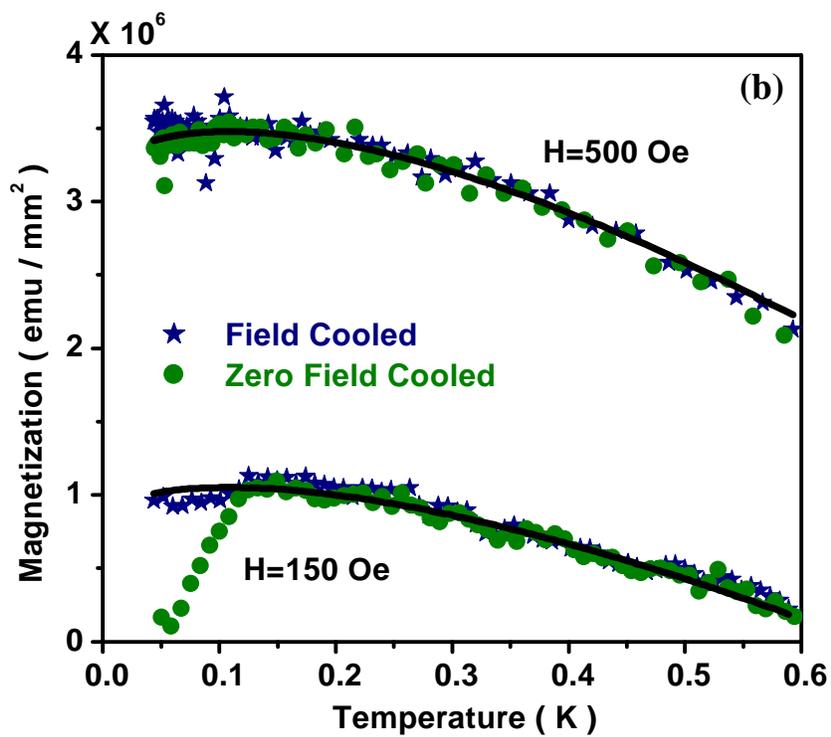
\end{document}